\documentclass[preprint]{aastex}
\usepackage{color, soul}

\slugcomment{}

\begin{document}


%\title{Detection of Radio Emission from the Super-massive Black Hole in M32}
\title{DETECTION OF A COMPACT NUCLEAR RADIO SOURCE IN THE LOCAL GROUP ELLIPTICAL GALAXY M32}

\author{Yang Yang\altaffilmark{1,2,3}, Zhiyuan Li\altaffilmark{1,2,3}, Lor\'{a}nt O. Sjouwerman\altaffilmark{4}, Q. Daniel Wang\altaffilmark{1,5}, Qiusheng Gu\altaffilmark{1,2,3}, Ralph P. Kraft\altaffilmark{6}, Feng Yuan\altaffilmark{7}}
\altaffiltext{1}{School of Astronomy and Space Science, Nanjing University, Nanjing 210093, China}
\altaffiltext{2}{Key Laboratory of Modern Astronomy and Astrophysics, Nanjing University, Nanjing 210093, China}
\altaffiltext{3}{Collaborative Innovation Center of Modern Astronomy and Space Exploration, Nanjing 210093, China}
\altaffiltext{4}{National Radio Astronomy Observatory, Socorro, NM 87801, USA}
\altaffiltext{5}{Department of Astronomy, University of Massachusetts, Amherst, MA 01003, USA}
\altaffiltext{6}{Harvard-Smithsonian Center for Astrophysics, 60 Garden Street, Cambridge, MA 02138, USA}
%\altaffiltext{7}{Department of Physics and Astronomy, University of California, Los Angeles, CA 90095}
\altaffiltext{7}{Shanghai Astronomical Observatory, Chinese Academy of Sciences, Shanghai 200030, China}

\email{lizy@nju.edu.cn}



\begin{abstract}
The Local Group compact elliptical galaxy M32 hosts one of the nearest candidate super-massive black holes (SMBHs), which has a previously suggested X-ray counterpart.
Based on sensitive observations taken with the {\it Karl G. Jansky} Very Large Array (VLA), we detect for the first time a compact radio source coincident with the nucleus of M32, 
which exhibits an integrated flux density of $\sim$$47.3\pm6.1$ $\mu$Jy at 6.6 GHz. We discuss several possibilities for the nature of this source, favoring an origin of the long-sought 
radio emission from the central SMBH, for which we also revisit the X-ray properties based on recently acquired {\sl Chandra} and {\sl XMM-Newton} data. Our VLA observations also discover radio emission from three previously known optical planetary nebulae in the inner region of  M32.
\end{abstract}

\section{Introduction}

It is recognized that super-massive black holes (SMBHs), commonly residing in the nuclei of present-day galaxies with a substantial stellar bulge, 
%Accretion onto and feedback (both radiative and mechanical) from the SMBH are among the fundamental astrophysical processes that govern galaxy evolution. 
have grown the bulk of their mass during quasar phases, but otherwise spent the majority of their life accreting at rates 
well below the Eddington limit (e.g., \citealp{1982MNRAS.200..115S,2002MNRAS.335..965Y,2004IAUS..222...49M}). 
As such, most SMBHs in the local Universe manifest themselves as low-luminosity active galactic nuclei (LLAGNs; see review by Ho 2008),
the imprint of which is often difficult to discern.
Nevertheless, dedicated surveys of nearby LLAGNs carried out in the past two decades have yielded high detection rates in the radio and X-ray bands (e.g., \citealp{2000ApJ...542..186N,2001ApJS..133...77H,2010ApJ...714...25G,2012ApJ...745L..13M}), providing important clues to their nature. 
The current consensus is that LLAGNs are powered by a radiatively inefficient, advection-dominated accretion flow (ADAF; \citealp{1994ApJ...428L..13N}) that operate at very sub-Eddington accretion rates, 
which is probably coupled with outflows in the form of jets and/or winds (see review by \citealt{2014ARA&A..52..529Y}). 
%Specific ADAF-jet models, involving jet synchrotron and Comptonization in the ADAF, have been confronted with the average
%broadband spectral energy distribution of individual LLAGNs, resulting in a varied degree of success (e.g., see review by Yuan 2007). 

In the Local Group, strong dynamical evidence of a SMBH have been found for all of the three galaxies with a prominent bulge: Milky Way, M31 and M32, among which our own Galaxy makes the most compelling case \citep{2010RvMP...82.3121G}.
Interestingly, these three nearest SMBHs also show the lowest Eddington ratios\footnote{Sgr A* and M31* have an Eddington ratio of $L_{\rm bol}/L_{\rm Edd} \sim10^{-8.5}$ (Baganoff et al.~2003; Li et al.~2011), where $L_{\rm bol}$ is the bolometric luminosity and $L_{\rm Edd}\equiv 1.3\times10^{38}[M_{\rm BH}/M_{\odot}]{\rm~erg~s^{-1}}$ the Eddington luminosity,
for masses of $\sim$$4\times10^6{\rm~M_\odot}$ \citep{Ghez08} and $\sim$$1.4\times10^8{\rm~M_\odot}$ (Bender et al.~2005), respectively. The present work infers $L_{\rm bol}/L_{\rm Edd} \sim10^{-7.5}$ for the SMBH in M32.} among known LLAGNs,
%This is a factor of $10^2-10^6$ lower than the inferred values for most other known LLAGNs (e.g., Ho 2009). 
thus supplying us with unique opportunities to explore the poorly understood physics of SMBHs at the most quiescent state.

In the compact elliptical galaxy M32, a central SMBH of $\sim$$2.5\times10^6{\rm~M_\odot}$ has been inferred from modeling of its circumnuclear stellar kinematics \citep{2002MNRAS.335..517V,2010MNRAS.401.1770V}. Additional evidence for an accreting SMBH comes from the {\it Chandra} detection of an X-ray source coincident with the nucleus of M32 (\citealt{2003ApJ...589..783H}, hereafter HTU03), which exhibits a power-law spectrum with a remarkably low 2-10 keV luminosity of $\sim$$10^{36}{\rm~erg~s^{-1}}$, corresponding to only $\sim$$10^{-8.5}$ of its Eddington luminosity. 
On the other hand, the SMBH leaves little, if any, signature at the optical and infrared bands, presumably owing to the dearth of interstellar medium (ISM) in M32 (e.g., \citealt{1997ApJS..112..315H,1998ApJ...507..726S,2001ApJ...557..671W,2007A&A...473..783R}; but see \citealt{2010ApJ...725..670S}). 
Even at the seemingly more promising radio bands, where synchrotron radiation from a putative jet is expected, a firm detection in the literature has been absent. Compiling Very Large Array observations taken before 2000, HTU03 derived 5\,$\sigma$ upper limits of 875, 30, 525 and 550 $\mu$Jy at 15, 8.4, 5 and 1.4 GHz, respectively.

In this work, we report the detection of a compact radio source at the nucleus of M32 based on high-resolution, sensitive observations recently carried out with the Karl G. Jansky Very Large Array (VLA). 
We also revisit the properties of the X-ray nucleus identified by HTU03, using recently acquired {\it Chandra} data {\it XMM-Newton} data. 
%Discussion of the results and a summary are presented in Section \ref{sec:discuss}. 
We adopt a distance of 780 kpc for M32, effectively placing it at the same distance as M31 \citep{2008ChJAA...8..369Y}. 
Throughout this work we quote 1$\sigma$ uncertainties unless otherwise stated. All coordinates are given for epoch J2000.

\section{Observations} \label{sec:data}

\subsection{Radio data} \label{subsec:rdata}

Three VLA observations of M32 were carried out in the B-configuration on July 24, 28 and 29, 2012 (Program ID: 12A-243; PI: L.~Sjouwerman).
Taken in the $C$-band, each observation has a total bandwidth of 2 GHz centered at 6.6 GHz and an integration time of 6 hrs.  
The flux calibrator was 3C48 with a 3\% systematic uncertainty \citep{2013ApJS..204...19P}.
%\footnote{http://www.aoc.nrao.edu/~smyers/calibration/2009/} 5.432 $\pm$ 0.027 Jy
The observations were phase-referenced to the calibrator J0038+4137 every 8 minutes with a  switching angle of 1$\fdg$1. The positional accuracy of our phase calibrator was $\lesssim$0$\farcs$002. 
%\footnote{http://www.vlba.nrao.edu/astro/calib/vlbaCalib.txt}. 
For each observation, the data was flagged, calibrated, and imaged following standard procedures with the Common Astronomy Software Applications\footnote{http://casa.nrao.edu} (CASA, version 3.4.0). 
We have used the SPLIT task to average the 64 channels of each spectral window with width = 2, and employed the CLEAN task using the Multi-Frequency Synthesis mode, nterms = 1, and Natural weighting to achieve maximum sensitivity.
The resultant images were made for a primary beam of $\sim$380$\arcsec$ and sampled the sysnthesized beam of 1$\farcs$14$\times$1$\farcs$09 with 0$\farcs$3 pixels. 
%The image shown in Figure~\ref{Fig:image}a shows the inner 25$\arcsec \times$25$\arcsec$. 
We have also used the CONCAT task to concatenate the three visibility data sets to produce the final, most sensitive image (Figure~\ref{Fig:image}a), 
which has an average rms noise of 1.2 $\mu$Jy beam$^{-1}$. Notably, this is a factor of $\sim$90 (5) lower than that achieved by HTU03 at the neighboring frequency of 5 (8.4) GHz. 
%A log of the VLA observations is given in Table \ref{tab:VLA}. 

\subsection{X-ray data} \label{subsec:xdata}
The field of M32 has been observed by {\it Chandra} with its Advanced CCD Imaging Spectrometer (ACIS) on nine epochs between 2000-2005. 
For each of these observations, the raw data was downloaded from the public archive and reprocessed following the standard procedure, using CIAO v4.3 and the corresponding calibration files.
%\footnote{http://cxc.harvard.edu/ciao/}. 
The relative astrometry among the individual observations was calibrated by matching centroids of point-like sources commonly 
detected in the overlapping field of view. 
In addition, we have carried out deep {\it XMM-Newton} observations of M32 (PI: Q.D.~Wang) on four epochs in 2011, with a total exposure of $\sim$400 ks.
For each observation, the data obtained from the European Photon Imaging Camera (EPIC) was reduced following 
the standard procedure, using SAS v11.0.1 and the corresponding calibration files.
%\footnote{http://heasarc.gsfc.nasa.gov/docs/xmm/abc/}. 
Table \ref{tab:chandra} gives a log of the X-ray data.

\section{Analysis and results} \label{sec:anal}
\subsection{Detection of radio sources} \label{subsec:radios}

Figure~\ref{Fig:image}a displays the inner 25$\arcsec$$\times$25$\arcsec$ region of M32 in the concatenated $C$-band image, overlaid by optical intensity contours (cyan) that trace the starlight. A compact source, which we designate R1, is clearly present at the optical center of M32.
R1 shows a peak flux density $\sim$10.7$\pm$1.2 $\mu$Jy beam$^{-1}$, but it appears more extended than a point source. Thus we performed a single elliptical Gaussian fit to R1 using the CASA task IMFIT. The fit suggested that the source is formally larger than the synthesized beam,
with a deconvolved FWHM $\approx$ [$(2\farcs8\pm0\farcs2)\times(1\farcs8\pm0\farcs2)$] ($\sim$10.6 pc $\times$ 6.8 pc) and a positional angle of 130$\fdg$1 $\pm$1$\fdg$9.
The peak of R1 was found at [RA, DEC]=[00$^{h}42^{m}41\fs838\pm0\fs007$, $+40\arcdeg51\arcmin54\farcs89\pm0\farcs07$], which is coincident with the 2MASS position of the nucleus, 
[00$^{h}$42$^{m}$41\fs825, +40\arcdeg51\arcmin54\farcs61], to within the uncertainty of the latter ($\lesssim$0\farcs5; \citealt{2003AJ....125..525J}).
An integrated flux density of 47.3$\pm$6.1 $\mu$Jy was measured for R1 in the concatenated image. The uncertainty is the quadratic sum of the error reported by IMFIT and the 3\% uncertainty in the absolute flux scale. 
%There is a slightly extension toward the northeast of the center. Indeed, the FWHM of Gaussian fitting result is more extensions than the FWHM of beam. It maybe imply that partial radio emission come from jet. 
We also measured the integrated flux density of R1 in the individual observations, but found no statistically significant variation. 
We further divided the full bandwidth into two 1-GHz sub-bands centering at 6 and 7 GHz to measure the spectral index, $\alpha$, defined as S$_{\nu}$ $\varpropto$ $\nu^{\alpha}$. The result was only poorly-constrained, with $\alpha \approx -2.0\pm1.7$.
%We conclude that the presence of a radio source at the nucleus of M32 is established. 

Three faint, off-nucleus sources are also evident in Figure~\ref{Fig:image}a, which we designate as R2, R3 and R4. 
They all lie within a projected distance of $10^{\prime\prime}$ from the nucleus and have a significance of $\sim$3\,$\sigma$ in the concatenated image. Table~\ref{tab:radiosource} gives their positions and flux densities.  
%[RA, DEC]=[ 00:42:41.891 $\pm$ 0.002,+40.51.58.412 $\pm$ 0.026],[00:42:42.563 $\pm$ 0.003,+40.51.56.552 $\pm$ 0.036],[00:42:42.659 $\pm$ 0.002,+40.51.56.465 $\pm$ 0.035]. 
Interestingly, all three sources are spatially coincident with planetary nebula (PNe) identified in the SAURON integral field spectroscopic 
observations of M32 (\citealt{Sarz11}; their PNe \#5, \#3 and \#7 are coincident with R2, R3 and R4, respectively). 
This is demonstrated by contrasting the white and cyan intensity contours in Figure~\ref{Fig:image}a. The latter are derived 
from an archival {\it Hubble Space Telescope} (HST) WFC/F502N image (Program ID: 11714).
% which we have collected from the Hubble Legacy Archive\footnote{http://hla.stsci.edu}. 
At the position of each off-nuclear radio source, a compact F502N source, presumably dominated by the [O III]$\lambda$5007 line emission, is clearly seen.
%Furthermore, R3 may also have a counterpart in the Slong Digital Sky Survey, which is named SDSS J104037.93+405155.7. 
Therefore, we suggest that the three off-nucleus sources are radio counterparts of known PNe in M32, which most likely arise from free-free emission (Kwok 2000). To our knowledge, this signifies the first detection of radio PNe at the far-side of the Local Group.
The 6.6 GHz monochromatic luminosities\footnote{L$_R\equiv\nu$L$_{\nu}$} of the three sources, $\sim$$2\times10^{31}{\rm~erg~s^{-1}}$,
are comparable to NGC\,7027, one of the most luminous Galactic PNe with a 4.9 GHz monochromatic luminosity of $\sim$$3\times10^{31}{\rm~erg~s^{-1}}$ \citep{Zijl08}.

\subsection{Revisiting the X-ray nucleus} \label{subsec:xrays}
Figure~\ref{Fig:image}b shows a 0.3-8 keV image for the same region as in Figure~\ref{Fig:image}a, obtained by combining the nine {\it Chandra}/ACIS observations. 
Three discrete sources, which have been previously designated as X1, X2 and X3 by HTU03, are clearly present, and they remain the only sources detected within
this region, despite the $\sim$3 times deeper total exposure achieved here. 
Among them, X1 is coincident with the nucleus and has been suggested to be the X-ray counterpart of the SMBH by HTU03. The slight offset ($\sim$0\farcs34) between the centriod of X1 and the nucleus in the WFC/F502N image is consistent with the astrometry accuracy\footnote{The typical pointing accuracy of both 
HST and {\sl Chandra} is $\lesssim1^{\prime\prime}$. 
However, owing to the lack of common sources, it is virtually impossible to match the optical/X-ray astrometry to better than the individual pointing accuracy.}.
From each ACIS observation, we extracted a spectrum for X1 from a 2$^{\prime\prime}$-radius circle, and a background spectrum from an annulus with 
inner-to-outer radii of $2^{\prime\prime}$--$4^{\prime\prime}$. In the following we restrict our analysis to energies above 0.5 keV, so that contamination from X2, the nearby very soft source, 
is negligible. We fitted the individual spectra using an absorbed power-law model ({\it wabs*powerlaw} in XSPEC), fixing the equivalent hydrogen column density at the Galactic foreground value, 
$7\times10^{20}{\rm~cm^{-2}}$. This is justified by the fact that M32 exhibits no detectable neutral ISM \citep{2001ApJ...557..671W}. For ObsIDs 2017, 2494 and 5690, which have good signal-to-noise ratios, we were able to constrain the photon-index, $\Gamma$, to be $2.50^{+0.17}_{-0.17}$, $2.72^{+0.21}_{-0.18}$ and $2.36^{+0.09}_{-0.10}$, respectively. 
For the remaining ObsIDs, we fixed the photon-index at 2.4 and constrained the normalization only. The spectral models were then used to predict
the 0.5-10 keV intrinsic luminosity of X1 at each epoch.

The moderate angular resolution afforded by the {\it XMM-Newton}/EPIC observations is not optimal for isolating the emission from X1, given the presence of the 
brighter source X3 located $\sim$$9^{\prime\prime}$ away (Figure~\ref{Fig:image}b). Therefore, we employed a two-dimensional image fitting method \citep{2011ApJ...728L..10L} to simultaneously 
determine the observed fluxes of X1 and X3, which not only maximizes the counting statistics, but also accounts for the propagation of the dominating error arising 
from the mutual PSF scattering. Briefly, we fitted the 0.5-8 keV image of the central $\sim$$60^{\prime\prime}\times60^{\prime\prime}$ field with a 
three-component model convolved with the PSF. The model consists of a constant local background and two point sources (for X1 and X3), each represented by a delta function. 
We mimicked the PSF using the image of a bright transient source recently found in the bulge of 
M31 (\citealt{2012AAS...21944205K}), averaged over four {\it XMM-Newton} observations (ObsIDs 0600660301, 0600660401, 0600660501 and 0600660601). 
We jointly fitted the EPIC MOS1 and MOS2 images taking into account their distinct PSFs.
We did not include data from the PN due to its poorer angular resolution. In the fit, we fixed the relative offset between X1 and X3, as determined from the combined {\it Chandra}/ACIS image.  
The best-fit was obtained by minimizing the C-statistic \citep{1979A&A....72L...6C}. The resultant count rates and uncertainties of X1 from the 
individual observations were then converted into an intrinsic luminosity by adopting the above absorbed power-law spectral model ($\Gamma=2.4$). 
We have added to the error budge a 10\% uncertainty in quadrature, to account for the uncertainty in the adopted photon-index. 
A consistency check indicates that the average flux ratio between X1 and X3 is in good agreement with that measured from the {\it Chandra} observations. 

The 0.5-10 keV intrinsic luminosity and its uncertainty thus determined for each observation are given in the last column of Table \ref{tab:chandra}.
Figure~\ref{Fig:vary} shows the long-term light curve of X1 over $\sim$11 yrs.
%There is a marginal trend of increasing luminosity, by a factor of $\sim$2 since 2000, for which we found a Spearman's rank correlation coefficient of 0.38 at $\sim$78\% confidence level. 
 The average luminosity seems to have increased by a factor of $\sim$2 since 2002, although the sparse temporal sampling could have missed intervals of substantially weaker or stronger radiation.
In addition, the luminosity was seen to increase by $\sim$$(190\pm100)$\% between two {\it XMM-Newton} observations separated by two days (Table \ref{tab:chandra}). 
We also searched for but found no significant intra-observation flux variation in either the {\it Chandra} or {\it XMM-Newton} exposures.

\section{Discussion and summary} \label{sec:discuss}

The VLA observations with unprecedented sensitivity have helped us establish the presence a compact radio source in the nucleus of M32. 
The source, R1, has a monochromatic luminosity of $\sim$$2.3\times10^{32}{\rm~erg~s^{-1}}$ at 6.6 GHz. 
This remarkably weak emission and the lack of significant variability hinder a straightforward identification of its nature.
We discuss several possibilities below and favor an association with the putative SMBH. 
%Following the nomenclature of Sgr A*, the radio counterpart of the Galactic Center Black Hole, hereafter we refer to this radio nuclear source in M32 as M32*. Its luminosity is less than 4 orders magnitude and its mass is same as the center source in our Galaxy, Sgr A*. 
%The central compact radio source (see Figure 1) is spatially coincident with a X-ray point source observed with $Chandra$ X-ray Observatory \citep{2003ApJ...589..783H}.
 %Following the nomenclature used in the Galaxy and M31, where Sgr A* and M31* are used to refer to the center SMBH, we hereafter refer to the potential radio nuclear counterpart in M32 as M32* later on. 

First, R1 can in principle arise from by synchrotron or free-free emission powered by, for instance, nuclear star-forming activities, in which case the source might span a physical size of $\sim$10 pc, as its slightly extended morphology suggests (Section~\ref{subsec:radios}).
However, there is virtually no evidence for recent or on-going star formation at the center of M32 -- in fact, none across the entire galaxy \citep{2000ApJ...532..308B}, disfavoring the presence of any recent core-collasped supernovae.
The expected rate of Type Ia supernovae at the nucleus is also very low ($\lesssim10^{-8}{\rm~yr^{-1}}$; HTU03). Thus, R1 is unlikely to be associated with synchrotron emission from a supernova remnant. Neither is there evidence for a circumnuclear ionized gas in M32. 
Ho et al.~(2003) reported a 3$\sigma$ upper limit in the H$\alpha$ luminosity of $\lesssim2\times10^{36}{\rm~erg~s^{-1}}$. Assuming a typical electron temperature $T_e \approx 10^4 \,\rm K$, we can use the following relation to estimate the strength of free-free emission from any circumnuclear ionized gas \citep{1986A&A...155..297C}: %report that if the $N(He^+)/N(H^+) \sim0.08$, the thermal flux density $S_T$ may be compared with H$\beta$ line flux $F({\rm H}\beta)$.
\begin{equation}
\left[\frac{F(\rm H\alpha)}{10^{-12} {\rm~erg\;cm^{-2}\,s^{-1}}}\right]\sim0.8\left(\frac{T_e}{10^4{\rm K}}\right)^{-0.52}\left(\frac{\nu}{{\rm GHz}}\right)^{0.1}\left(\frac{S_{{\rm ff}}}{\rm mJy}\right).
\end{equation}
The above upper limit in H$\alpha$ luminosity corresponds to $S_{\rm ff,\, 6.6\,{\rm GHz}}$ $<$ 28.4 $\mu$Jy. This is significantly lower than the flux density of R1, suggesting that the bulk of the observed radio emission does not come from free-free emission.
  
Alternatively, we might have caught a transient radio source similar to those found in the center of our Galaxy (e.g., \citealt{1976MNRAS.177..319D,1992Sci...255.1538Z,2002AJ....123.1497H,2009ApJ...696..280H}). In particular, 
the Galactic Center Transient (GCT) detected at a projected distance of $\sim$1.4 pc from Sgr A* during 1990-1991 would have a 6.6 GHz flux density of $\sim$10-20 $\mu$Jy if located at the distance of M32 (Zhao et al.~1992). 
The nature of some of these transients, including the GCT, remains uncertain. One favored suggestion was that they are radio outbursts of black hole binaries (BHBs), although in most cases the expected X-ray emission was actually not detected.
R1 could be such a radio transient, provided that it was not physically related to the stable source X1.

Can, then, both R1 and X1 be the persistent counterpart of a BHB located in the center of M32? Indeed, the identification of X1 as the X-ray counterpart of the SMBH essentially rested on positional coincidence (HTU03). The very modest luminosity of X1, $\sim$$10^{36}{\rm~erg~s^{-1}}$, 
lies within the range of X-ray binaries (XRBs) in the low/hard state.                           
The {\it Chandra} and {\it XMM-Newton} observations now suggest that X1 has persisted over a decade (Section~\ref{subsec:xrays}), hence it is not an X-ray transient similar to those found in the Galactic center, which are likely outbursting BHBs \citep{2005ApJ...622L.113M}.
Moreover, its steep power-law spectrum ($\Gamma$$\sim$2.4) is atypical of XRBs in the low/hard state, which generally exhibit $\Gamma$ $\sim$ 1.5--2 (e.g., \citealt{2006ARA&A..44...49R}, \citealt{2008ApJ...682..212W}). Furthermore,
the BHB case is strongly disfavored on the basis of the ``fundamental plane of black hole activity" (hereafter the FP; \citealt{2003MNRAS.345.1057M,2004A&A...414..895F}), an empirical relation linking the radio luminosity ($L_{\rm R}$, practically evaluated at 5 GHz), X-ray luminosity ($L_{\rm X}$, evaluated as $L_{2-10\,\rm keV}$) and black hole mass ($M_{\rm BH}$), in the form of \cite{2009ApJ...706..404G}:
\begin{equation}
% logL_R = 0.60^{+0.11}_{-0.11}logL_X + 0.78^{+0.11}_{-0.09}logM_{\rm BH} + 7.33^{+4.05}_{-4.07}.
 {\rm log}\, L_{\rm R} = (4.80 \pm 0.24) + (0.78 \pm 0.27)\,{\rm log}\,M_{\rm BH} + (0.67 \pm 0.12)\,{\rm log}\, L_{\rm X}
 \end{equation}
Following this relation, a BHB with $M_{\rm BH}$ $\sim10 M_{\sun}$ and  $L_{2-10\, \rm keV}\footnote{Scaled from the average 0.5-10 keV luminosity assuming the fiducial power-law spectrum (Section~\ref{subsec:xrays}).}\sim10^{36}$ erg s$^{-1}$, is expected to have $L_{5\, \rm GHz}$ of
$\sim5\times10^{29}$ erg s$^{-1}$. This is $\sim$400 times lower than our measurement for R1, when extrapolated from the neighboring frequency of 6.6 GHz assuming a canonical spectral index of $\alpha = -0.7$.
%Assuming L$_{5GHz}$ = $\nu$L$_{\nu}$ $\approx$ 3.3$\pm$0.4 $\times$ 10$^{31}$ erg s$^{-1}$ with the distance of M32. The ratio of radio to 2-10 keV luminosity, 
%log(L$_{5GHz}$/L$_X$)$\approx$ -4.5, where radio-loud is defined as log(L$_{5GHz}$/L$_X$) $>$ -4.5 \citep{2003ApJ...583..145T}. 
%This is consistent with the typically radio-loud nature of LLAGN in general \citep{2008ARA&A..46..475H}. 
%Based on \cite{2003MNRAS.345.1057M} shows the "Fundamental Plane of Black hole Activity", the FP relationship is defined by:

Lastly, R1 may be the long-sought radio counterpart of the central SMBH, which most likely arises from radio synchrotron radiation. If this were the case, 
R1 would be one of the weakest known ``AGNs'' in radio, along with Sgr A* ($L_{5\,\rm GHz} \sim 3\times 10^{32} $ erg s$^{-1}$; \citealt{2001ApJ...547L..29Z}) and M31* ($L_{5\, \rm GHz}\sim 2\times 10^{32}$ erg s$^{-1}$; \citealt{2010ApJ...710..755G}). 
%\textbf{On the other hand, the X-ray luminosities of Sgr A* and M31* have been found to vary by a factor of $10^{2}-10^{3}$ on short timescales \citep{2001Natur.413...45B,2011ApJ...728L..10L}.} 
%To further relate the radio and X-ray properties of the SMBH in M32, let us first revisit the likelihood of X1 being a genuine LLAGN.
On the other hand, the measured value of $\Gamma$ ($\sim$2.4) for X1 would be compatible with 
the empirical anti-correlation between the photon-index and the Eddington ratio in LLAGNs \citep{2009MNRAS.399..349G}, for $L_{\rm bol}/L_{\rm Edd}\sim10^{-7.5}$ of the SMBH in M32.
%In such a case, the X-ray emission is likely from a synchrotron-cooled jet (Heinz 2004). 
This supports X1 being the genuine X-ray counterpart of the SMBH.
% The moderate flux variability in X1 (Section~\ref{subsec:xrays}) is not atypical of LLAGNs, while the order-unity short-term variability appears small compared to that found in Sgr A* and M31* (up to a factor of $10^{2}$; \citealp{2001Natur.413...45B,2011ApJ...728L..10L}).

Now, assuming that both R1 and X1 are the manifestations of the central SMBH, it would be interesting to confront the radio and X-ray measurements with the FP. The ``classical'' FP (Equation 2) predicts $L_{\rm R} \sim 8.1\times10^{33}$ erg s$^{-1}$ (for $M_{\rm BH}\approx2.5\times10^6$ M$_{\sun}$ and $L_{\rm X} \approx 1.0\times10^{36}$ erg s$^{-1}$).
%\begin{equation}
 %log L_R = (0.60^{+ 0.11}_{-0.11}) log L_X+(0.78^{+ 0.11}_{-0.09}) logM_{\rm BH} +7.33^{+ 4.05}_{-4.07} 
 %\end{equation}
Alternatively, the modified FP proposed by Yuan \& Cui (2005) and Yuan et al. (2009; {Equation 5 therein),
\begin{equation}
 {\rm log}\,L_{\rm R} = 1.29(\pm 0.03)\,{\rm log}\, L_{\rm X}+0.11(\pm 0.04)\,{\rm log}\,M_{\rm BH} -14.1, 
 \end{equation}
which should operate at the low-luminosity regime ($L_{\rm X}\lesssim10^{-6}L_{\rm Edd}$) where the X-ray emission might be dominated by a synchrotron-cooled jet, predicts $L_{\rm R}$ $\sim$ 1.1$\times10^{33}$ erg s$^{-1}$.
Taking into account the scatter ($\sim$1.0 dex; \citealt{2009ApJ...706..404G}) in the FPs, the predicted values are not inconsistent with the observed luminosity ($2.3\times 10^{32}$ erg s$^{-1}$).
This can be further considered a supporting evidence that both R1 and X1 are indeed physically related to the central SMBH. 
In this case, the slightly extended morphology of R1 might trace pc-scale jets/lobes or a core-plus-jets system. 
%while the substanial short-term variability in X1 might reflect rapid changes in the jet energetics.

In summary, our sensitive VLA observations have revealed a weak radio source coincident with the nucleus of M32. While at this stage a stellar
origin cannot be completely ruled out for this nuclear source, we favor the interpretation that it is the long-sought radio counterpart of the central SMBH, and suggest to designate it M32* 
for future reference, following the convention of Sgr A*, the radio counterpart of the Galactic Center Black Hole. Future VLA observations in multi-wavelength and with higher angular resolutions will be useful to unambiguously determine the nature 
of this source. Such observations will also benefit an in-depth investigation of the serendipitously detected radio PNe in the vicinity of M32*.

\acknowledgments
The National Radio Astronomy Observatory is a facility of the National Science Foundation operated under cooperative agreement by Associated Universities, Inc. We thank M. Sarzi for valuable comments on the SAURON observations of PNe in M32. This work was partially supported by the National Natural Science Foundation of China (grants 11473010 and 11133001). 
Y.Y. is indebted to the hospitality of NRAO during her visit. Z.L. acknowledges support from the Recruitment Program of Global Youth Experts.

\begin{figure*}\centering
\epsscale{1}
\plotone{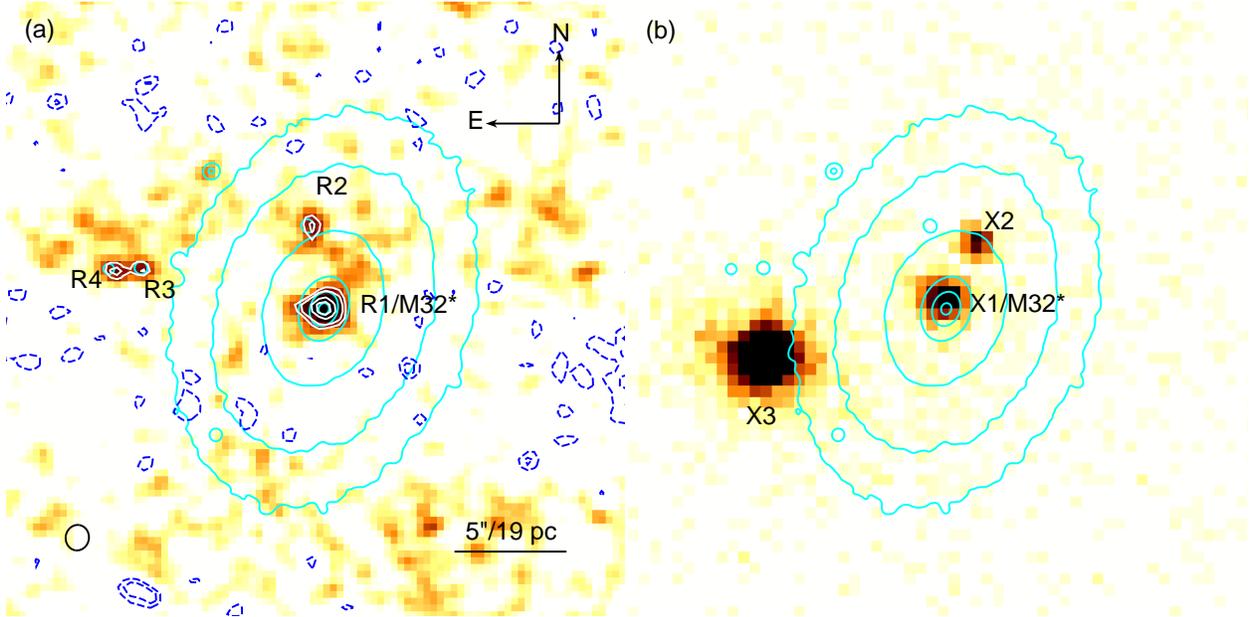}
\caption{(a) The 18-hr integrated VLA $C$-band image of the central 25$\arcsec$$\times$ 25$\arcsec$ ($\sim$95 pc $\times$ 95 pc) region of M32, overlaid by intensity contours (positive in white solid and negative in blue dashed)
at levels of $-$3, $-$2, 4, 5, 7 and 8 times the average rms (1.2 $\mu$Jy beam$^{-1}$). Also shown are HST/WFC3/F502N intensity contours (cyan) highlighting the starlight distribution of M32. Four compact sources are labeled, among which R1 is coincident with the nucleus of M32 (named M32*), while R2, R3 and R4 are coincident with optical PNe. 
The black ellipse on the lower left-hand corner represents the synthesized beam size.
(b) A {\it Chandra} 0.3-8 keV image showing the same region, overlaid by the F502N intensity contours. The three X-ray sources are labeled, following the convention of HTU03.}
\label{Fig:image}
\end{figure*}

\begin{figure*}\centering
\includegraphics[scale=0.9,angle=90]{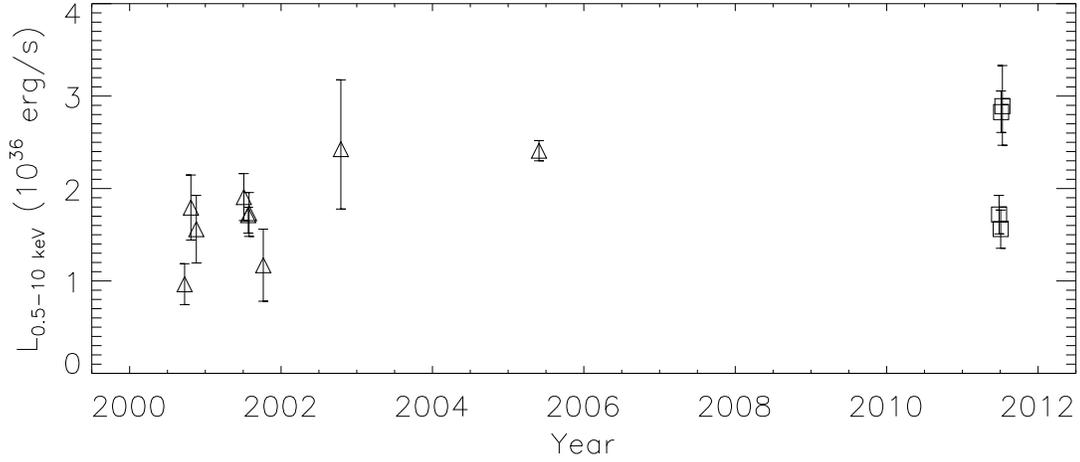}
\caption{A long-term 0.5-10 keV light curve of X1 based on nine {\it Chandra} (triangles) and four {\it XMM-Newton} (squares) observations.
%The 0.5-10 keV intrinsic luminosities are derived from spectral fits (for {\it Chandra} observations with sufficient statistics) or converted from the observed count rates 
%using a fiducial spectral model (for all {\it XMM-Newton} observations and shallow {\it Chandra} observations). 
See text for details.}
\label{Fig:vary}
\end{figure*}

%\clearpage 

%\begin{deluxetable}{ccccccc}
%\tablecaption{{VLA} observations of M32}
%\tablewidth{0pt}
%\tablehead{
%\colhead{Date} &
%\colhead{Exposure} &
%\colhead{$\theta_{\rm maj}$} &
%\colhead{$\theta_{\rm min}$} &
%\colhead{$\theta_{\rm position}$} &
%\colhead{$I_{\rm p}$} &
%\colhead{$I_{\rm rms}$}\\
%\colhead{(1)} &
%\colhead{(2)} &
%\colhead{(3)} &
%\colhead{(4)} &
%\colhead{(5)} &
%\colhead{(6)} &
%\colhead{(7)} 
%}
%\startdata
% 2012-07-24 & 6.0 &  1.08   &1.03 &41.4& 12.6&1.9\\
%2012-07-28 & 6.0 &  1.21   &1.09 &-86.1.1& 10.8 &1.8\\
% 2012-07-29 & 6.0 &  1.08   &1.06 &78.9& 10.1&2.2
%\enddata
%\tablecomments{(1) Date of observation; (2) Integration time, in hours;  (3)-(4): Synthesized beam size of the major and minor axes, in units of arcsec; (5): Position angle of synthesized beam, in units of deg; (6) Peak intensity, in units of $\mu$Jy beam$^{-1}$;(7) RMS image noise level, in units of $\mu$Jy beam$^{-1}$.}
%\label{tab:VLA}
%\end{deluxetable}

%\begin{deluxetable}{cccccc}
\begin{deluxetable}{cccc}
\tabletypesize{\footnotesize}
\tablecaption{X-ray observations of M32}
\tablewidth{0pt}
\tablehead{
\colhead{ObsID} &
\colhead{Date} &
\colhead{Exposure} &
%\colhead{RA} &
%\colhead{DEC} &
\colhead{$L_{\rm X}$} \\
\colhead{(1)} &
\colhead{(2)} &
\colhead{(3)} &
%\colhead{(4)} &
%\colhead{(5)} &
\colhead{(6)} 
}
\startdata
%  313 (C/ACIS-S) & 2000-09-21&   6.0&   10.65816&   40.87837&   $1.0^{+0.2}_{-0.2}$\\
%  314 (C/ACIS-S)& 2000-10-21&   5.1&   10.64900&   40.86757&  $1.8^{+0.4}_{-0.4}$\\
% 1580 (C/ACIS-I)& 2000-11-17&   5.1&   10.66379&   40.85630 &  $1.6^{+0.4}_{-0.4}$ \\
% 1584 (C/ACIS-I)& 2001-07-03&   4.9&   10.66362&   40.92284 &  $1.9^{+0.3}_{-0.3}$ \\
% 2017 (C/ACIS-S)& 2001-07-24&  45.9&   10.66159&   40.88299 &  $1.7^{+0.2}_{-0.2}$ \\
% 2494 (C/ACIS-S)& 2001-07-28&  16.0&   10.66160&   40.88298 &  $1.7^{+0.3}_{-0.3}$\\
% 1576 (C/ACIS-I)& 2001-10-05&   4.9&   10.62845&   40.96574 &  $1.2^{+0.4}_{-0.4}$\\
% 2894 (C/ACIS-I)& 2002-10-14&   4.7&   10.63939&   40.95643 &  $2.4^{+0.7}_{-0.8}$\\
% 5690 (C/ACIS-S)& 2005-05-27& 113.0&   10.67788&   40.87440 &  $2.4^{+0.3}_{-0.3}$ \\
%  0672130101 (X) & 2011-06-27 &   99.4&   10.67308&   40.86722 & $1.9^{+0.3}_{-0.3}$ \\
%  0672130601 (X) & 2011-07-05 &   119.9&   10.67267&   40.86705 & $1.7^{+0.3}_{-0.3}$ \\
%  0672130701 (X) & 2011-07-07 &  101.7&   10.67354&   40.86703 & $3.2^{+0.4}_{-0.4}$  \\
%  0672130501 (X) & 2011-07-13 &  55.8&  10.67342 & 40.86626 & $2.8^{+0.4}_{-0.4}$
  313 (C/ACIS-S) & 2000-09-21&   6.0&   $1.0^{+0.2}_{-0.2}$\\
  314 (C/ACIS-S)& 2000-10-21&   5.1&   $1.8^{+0.4}_{-0.4}$\\
 1580 (C/ACIS-S)& 2000-11-17&   5.1&  $1.6^{+0.4}_{-0.4}$ \\
 1584 (C/ACIS-I)& 2001-07-03&   4.9&   $1.9^{+0.3}_{-0.3}$ \\
 2017 (C/ACIS-S)& 2001-07-24&  45.9&   $1.7^{+0.2}_{-0.2}$ \\
 2494 (C/ACIS-S)& 2001-07-28&  16.0&    $1.7^{+0.3}_{-0.3}$\\
 1576 (C/ACIS-I)& 2001-10-05&   4.9&   $1.2^{+0.4}_{-0.4}$\\
 2894 (C/ACIS-I)& 2002-10-14&   4.7&   $2.4^{+0.7}_{-0.8}$\\
 5690 (C/ACIS-S)& 2005-05-27& 113.0&    $2.4^{+0.3}_{-0.3}$ \\
  0672130101 (X) & 2011-06-27 &   99.4&   $1.9^{+0.3}_{-0.3}$ \\
  0672130601 (X) & 2011-07-05 &   119.9&  $1.7^{+0.3}_{-0.3}$ \\
  0672130701 (X) & 2011-07-07 &  101.7&  $3.2^{+0.4}_{-0.4}$  \\
  0672130501 (X) & 2011-07-13 &  55.8& $2.8^{+0.4}_{-0.4}$
\enddata
\tablecomments{(1) Observation ID. {\it Chandra} observations are denoted by ``C/ACIS-S'' or ``C/ACIS-I'' and {\it XMM-Newton} observations by ``X''; (2) Date of observation; (3) Effective exposure, in ks; 
%(4)-(5): Right ascension and Declination (J2000);
  (6) 0.5-10 keV unabsorbed luminosity of the nuclear source, in units of $10^{36}{\rm~erg~s^{-1}}$.}
\label{tab:chandra}
\end{deluxetable}

\begin{deluxetable}{cccccccccccc}
\tabletypesize{\footnotesize}
\tablecaption{Properties of detected radio sources}
\tablewidth{0pt}
\tablehead{
\colhead{Source} &
\colhead{RA} &
\colhead{DEC} &
\colhead{$I_{\rm p}$}&
\colhead{$S_{\rm 6.6}$} &
\colhead{$L_{\rm 6.6}$} &
\colhead{$L_{\rm [O III]}$} \\
\colhead{(1)} &
\colhead{(2)} &
\colhead{(3)} &
\colhead{(4)} &
\colhead{(5)} &
\colhead{(6)}&
\colhead{(7)} 
}
\startdata
 R1 & 00$^{h}$42$^{m}$41\fs838 & +40\arcdeg51\arcmin54\farcs98  &10.7$\pm$1.2& 47.3$\pm$6.1 & 22.7 &$<4.3$ \\
 R2 & 00$^{h}$42$^{m}$41\fs891 & +40\arcdeg51\arcmin58\farcs41  &6.6$\pm$1.2&3.5$\pm$1.2  & 1.7  & 11.1 &\\
 R3 & 00$^{h}$42$^{m}$42\fs563 & +40\arcdeg51\arcmin56\farcs52  &6.6$\pm$1.2&3.4$\pm$1.2  & 1.6  & 7.0 \\
 R4 & 00$^{h}$42$^{m}$42\fs659 & +40\arcdeg51\arcmin56\farcs41  &6.8$\pm$1.2&3.5$\pm$1.2 &1.7  & 5.6
\enddata
\tablecomments{(1) Source name; (2)-(3) Source position in Celestial coordinates; (4) Peak flux density, in units of $\mu$Jy beam$^{-1}$; (5) Total integrated flux density and error at 6.6 GHz, 
in units of $\mu$Jy;
(6) Monochromatic luminosity at 6.6 GHz, in units of $10^{31}{\rm~erg~s^{-1}}$; (7) [O III]$\lambda$5007 emission line luminosity (or 3\,$\sigma$ upper limit), 
in units of $10^{35}{\rm~erg~s^{-1}}$, from Sarzi et al.~(2011).}
\label{tab:radiosource}
\end{deluxetable}

\end{document}